\documentclass[%
 aip,
 amsmath,amssymb,
 reprint,%
]{revtex4-1}

\usepackage{graphicx}% Include figure files
\usepackage{dcolumn}% Align table columns on decimal point
\usepackage{bm}% bold math
\usepackage[utf8]{inputenc}
\usepackage[T1]{fontenc}
\usepackage{mathptmx}

\begin{document}

\preprint{AIP/123-QED}

\title[]{Membrane sandwich squeeze film pressure sensors}

\author{Aur\'{e}lien Dantan}
\email{dantan@phys.au.dk}
\affiliation{Department of Physics and Astronomy, University of Aarhus, DK-8000 Aarhus C, Denmark}

\date{\today}

\begin{abstract}
Squeeze film pressure sensors, exploiting the dynamical modification of the mechanical properties of oscillating elements due to the compression of a fluid in a small gap region, allow for direct and absolute pressure measurements. This tutorial article discusses the working principles of membrane sandwich squeeze film pressure sensors, i.e. sensors consisting in a parallel arrangement of two large area, ultrathin suspended films forming a few-micron gap and immersed in a fluid, and focuses on their operation in the free molecular flow regime. The effects of gas pressure on the vibrations of the membrane resonators and their coupled dynamics are discussed in general terms before recent experimental implementations using high tensile stress silicon nitride membranes are presented.
\end{abstract}

\maketitle

\section{Introduction}

Large area, low thickness drums are commonly used for pressure sensing. In conventional miniaturized capacitive or piezo-resistive sensors the pressure difference with a reference cavity is determined by measuring the deflection of the suspended drum or by dynamically monitoring the change of the drum resonance frequencies~\cite{Bhat2007,Bao2000}. Typically, capacitive and piezo-resistive sensors operate around atmospheric pressure or in low vacuum (millibar). For high and ultrahigh vacuum pressure measurements, ionization gauge sensors are routinely used due to their very high sensitivity, but they intrinsically require the knowledge of the involved species and their ionization cross-section in order to determine an absolute pressure.

Squeeze film pressure sensors---which exploit the dynamical modification of the mechanical properties of oscillating elements due to the compression of a fluid in a small gap region~\cite{Bao2007}---offer a solution to this problem. By measuring the compressive force, rather than the force due the pressure difference with a reference cavity, such sensors allow for direct and absolute pressure measurements and bypasses the challenge of fabricating stable, sealed and non-outgassing reference cavities. Displacement measurements allowing for determining the mechanical resonance frequencies and linewidths of the drums can be conveniently performed electrically in MEMs systems, although they require the use of electrodes/lossy materials, which may deteriorate the mechanical properties of the resonators and electronic noise ultimately limits the achievable sensitivity. Optical interferometric displacement measurements may in contrast be more sensitive and allow for the use of purely dielectric thin films, such as e.g. silicon nitride, with very low mass and excellent mechanical properties. 

Since Blech's proposal to use squeeze film effects to control the response of seismic interferometers~\cite{Blech1983}, squeeze film sensors have been realized with various materials, geometries and detection techniques~\cite{Prak1991,Blom1992,Andrews1993,Legtenberg1994,Veijola1995,Steeneken2004,Vignola2006,Verbridge2008,Mol2009,Suijlen2009,Southworth2009,Stifter2012,Suijlen2012,Kainz2014,Kumar2015,Dolleman2016,Naesby2017,Naserbakht2019sna}
Progress in fabricating and controlling micro- and nanostructures has in particular allowed for the application of high-frequency and high-mechanical quality resonators for the exploration of a broad range of hydrodynamics regime~\cite{Ekinci2005,Li2007,Svitelskiy2009,Ekinci2010,Arlett2011}.

This tutorial article focuses on the working principles of squeeze film pressure sensors based on "membrane sandwiches", i.e. pressure sensors consisting in a parallel arrangement of two large area, ultrathin suspended films forming a few-micron gap and immersed in a fluid. In the first part, the various pressure regimes and relevant characteristic numbers are introduced in order to discuss the nature of the interaction of the fluid with the oscillating membranes. This article focuses mostly on the free molecular flow regime, in which the fluid can be considered a gas which is isothermally compressed in the gap region. The basic equations describing the dynamics of the membrane vibrations are derived and the squeeze film effect consequences, in particular the gas-induced coupling between the membranes, are generally discussed. In the second part of the article, recent experimental implementations~\cite{Naesby2017,Naserbakht2019sna} and illustrations of these effects using high-tensile stress silicon nitride membrane resonators are presented.

\section{Squeeze film effects for oscillating plates}

\subsection{Free molecular flow {\it vs} viscous regime}\label{sec:free}

\begin{figure}
\includegraphics[width=0.8\columnwidth]{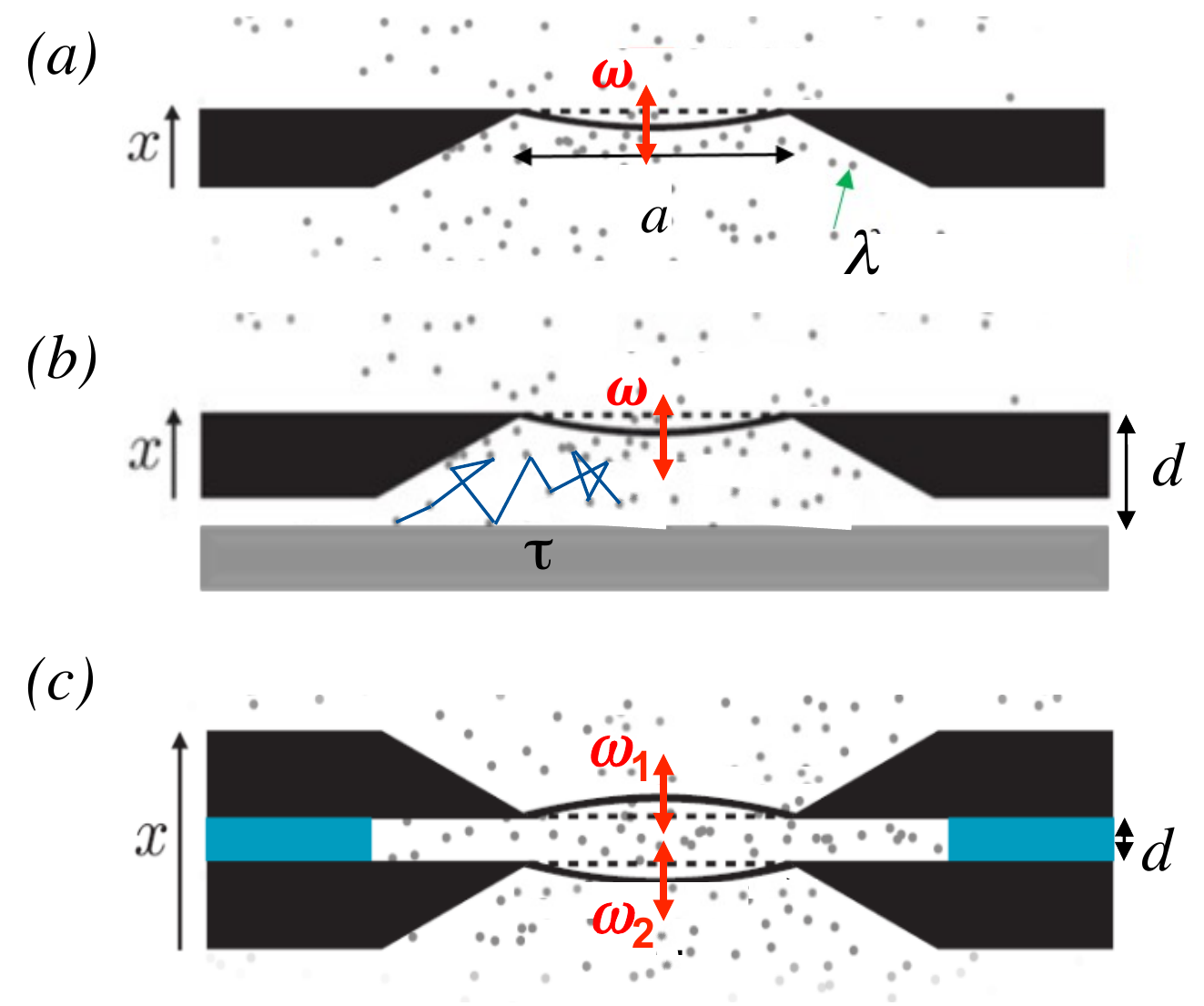}
\caption{(a) Single membrane oscillating in a fluid at pressure $P$. $\lambda$ is the molecule mean free path and $a$ the membrane's transverse dimension. (b) Single membrane oscillating in a fluid close to a fixed plate (gap $d$). $\tau$ is the average time that it takes for the molecules to leave the gap. (c) Membrane sandwich consisting of two parallel membranes oscillating at $\omega_1$ and $\omega_2$ and separated by a gap $d$.}
\label{fig:membranes}
\end{figure}

\begin{figure}
\includegraphics[width=\columnwidth]{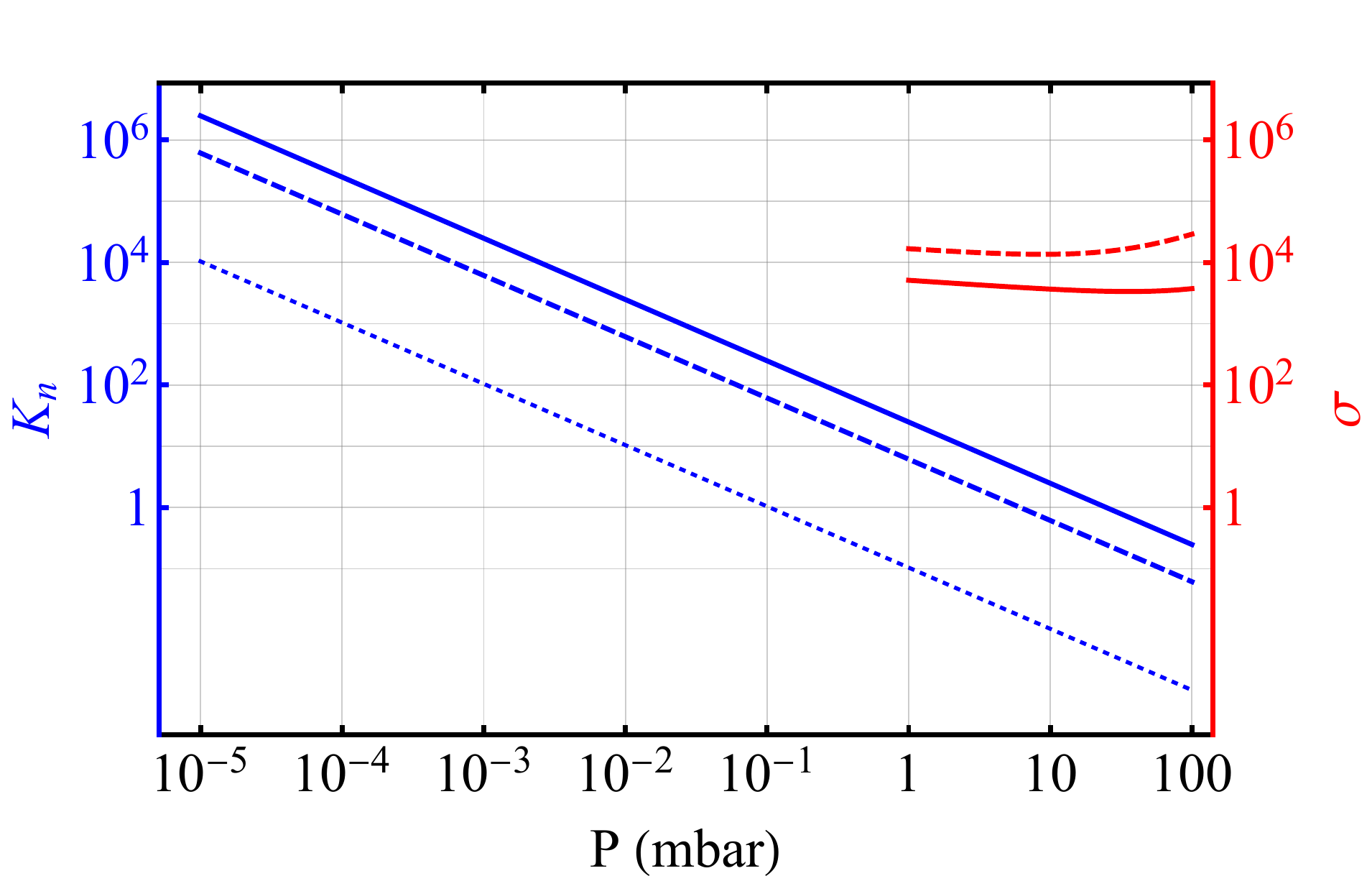}
%\llap{\parbox[b]{6.7in}{\large{{\it (a)}}\\\rule{0ex}{1.8in}}}
%\includegraphics[width=0.99\columnwidth]{knudsen_dampings2.pdf}
\caption{Variation of the Knudsen (blue) and squeeze (red) numbers with air pressure ($M_0=29$ g/mol, $\sigma_0=4.19\times 10^{-10}$ m), at room temperature and for a 500 $\mu$m-square, 100 nm-thick SiN membrane ($\rho=2700$ kg/m$^3$) and a gap of $d=2.1$ $\mu$m (plain) and $d=8.5$ $\mu$m (dashed). The blue dotted curve shows the evolution of the Knudsen number for the membrane far from any structure. Figure adapted from~\cite{Naserbakht2019sna}.}
\label{fig:knudsen}
\end{figure}

The nature of the interaction of the fluid particles with a moving plate, as depicted on Fig.~\ref{fig:membranes}(a), depends on the {\it Knudsen number}, defined as the ratio of the particle mean free path $\lambda$ to the typical plate dimension $a$,
\begin{equation}
K_n=\frac{\lambda}{a}=\frac{k_BT}{\sqrt{2}\pi\sigma_0^2Pa},
\end{equation}
where $k_B$ is the Boltzmann constant, $T$ the temperature, $\sigma_0$ the fluid molecule diameter and $P$ the pressure. At high Knudsen numbers ($K_n\gtrsim 10$), collisions between the molecules are negligible with respect to collisions with the plate, and the fluid can be considered behaving as a gas. This regime is thus commonly refered to as the free molecular flow (or rarefied air) regime. In contrast, at low Knudsen numbers ($K_n\lesssim 0.1$), the strong interactions between the molecules prompt for a viscous fluid description, involving the fluid viscosity $\mu$. The intermediate regime $0.1\lesssim K_n\lesssim 10$, in which neither gas or viscous fluid descriptions are legitimate, is then refered to as the transitional regime. Figure~\ref{fig:knudsen}(a) shows the evolution of the Knudsen number as a function of air pressure and at room temperature for a 0.5 mm-square plate, for which the transitional regime is in the millibar range.

In the free molecular flow regime the collisions of the gas molecules with the oscillating plate give rise to a viscous damping force, resulting in an additional, {\it kinetic} damping rate for the oscillations of the plate given by~\cite{Christian1966}
\begin{equation}
\gamma_\textrm{kin}=4\sqrt{\frac{2}{\pi}}\sqrt{\frac{M_0}{RT}}\frac{P}{\rho t},
\label{eq:gamma_kin}
\end{equation}
where $M_0$ is the gas molar mass, $R=8.31$ J/K/mol the ideal gas constant, $\rho$ the plate density and $t$ its thickness.

\subsection{Squeeze film force on an oscillating plate}

The previous analysis holds for a plate oscillating in a fluid far from other structures/boundaries. However, in presence of a nearby structure, e.g. a fixed parallel plate defining a small gap $d$ as illustrated in Fig.~\ref{fig:membranes}(b), the dynamics of the molecules inside the gap may be substantially modified~\cite{Bao2007}. The compression of the fluid inside the gap results in an additional squeeze film force on the moving plate. 

In the viscous regime, the relevant parameter to characterize the behavior of the compressed fluid is the {\it squeeze number} parameter
\begin{equation}
\sigma=\frac{\pi d^2P}{24a^2\mu},
\end{equation}
where $\mu$ is the fluid viscosity. The squeezed film dynamics in the viscous regime have been discussed in the literature~\cite{Blech1983,Veijola1995,Bao2007,Ekinci2010}, a high squeeze number giving rise to a predominantly elastic squeeze film force, whereas a low squeeze number leads to a predominantly dissipative force on the plate. As the pressure decreases, the squeeze number evolves in a non-trivial way due to the pressure dependence of the viscosity, which can to some extent be estimated phenomelogically~\cite{Veijola1995}.

We focus here on the free molecular flow regime, in which the relevant quantity to discuss the nature of the trapping of the fluid is the {\it Weissenberg number}~\cite{Ekinci2010}, 
\begin{equation}
W_i=\omega\tau,
\end{equation}
i.e. the product of the mechanical frequency $\omega$ by the molecular diffusion time $\tau$, which
represents the time constant for equalizing the pressure inside the gap.

The relaxation time corresponds to the average time that it takes for the molecules to leave the
gap. For a large plate ($a\gg d$) and assuming essentially inelastic collisions with the plates the relaxation time can be estimated by considering that the molecules diffuse inside the gap according to a random walk~\cite{Suijlen2009}, yielding
\begin{equation}
\tau=\frac{8a^2}{\pi^3d\bar{v}},
\label{eq:relax}
\end{equation}
where $\bar{v}=\sqrt{8RT/\pi M_0}$ is the mean velocity of the gas molecules. For elastic or partially elastic collisions, Bao's energy transfer model~\cite{Bao2002} or molecular dynamics simulations~\cite{Hutcherson2004,Bao2007} may alternatively be used to estimate $\tau$.

Following Suijlen {\it et al.}~\cite{Suijlen2009}, the squeeze film force can be related to the variation of the pressure inside the gap $\Delta P(t)$, which, in the case of an isothermal compression, is related to the density variation $\Delta n(t)$ inside the gap 
\begin{equation}
F_\textrm{sq}=\Delta P(t)A=\Delta n(t) k_BTA,
\end{equation}
where $A=a^2$ is the plate area. The relative density variation depends, on the one hand, on the random walk diffusion of the molecules and, on the other hand, on the volume change due to the displacement of the plate. It thus obeys the differential equation
\begin{equation}
\frac{d}{dt}\left(\frac{\Delta n}{n}\right)=-\frac{1}{\tau}\frac{\Delta n}{n}-\frac{d}{dt}\left(\frac{x}{d}\right),
\end{equation}
where $n$ is the equilibrium density and $x$ the moving plate displacement with respect to its equilibrium position, which we arbitrarily take positive for an increasing gap. For forced oscillations at frequency $\omega$, the complex amplitudes of the density variation $\Delta \tilde{n}$ and the displacement $\tilde{x}$ are then related by
\begin{equation}
\frac{\Delta \tilde{n}}{n}=-\frac{\tilde{x}}{d}\frac{i\omega\tau}{1+i\omega\tau}.
\label{eq:Deltan}
\end{equation}
The complex amplitude of the squeeze film force on the plate is then given by
\begin{equation}
\tilde{F}_\textrm{sq}=-\frac{PA}{d}\frac{i\omega\tau}{1+i\omega\tau}\tilde{x}.
\end{equation}
The real part of the squeeze film force, opposite in phase with the plate motion, contributes to an additional stiffness
\begin{equation}
k_\textrm{sq}=\frac{PA}{d}\frac{(\omega\tau)^2}{1+(\omega\tau)^2}
\label{eq:k_sq}
\end{equation}
to the plate's inherent stiffness $k_m$, while its imaginary part, in phase opposition with the plate velocity, contributes to an additional viscous damping rate
\begin{equation}
\gamma_\textrm{sq}=\frac{P\tau}{\rho t d}\frac{1}{1+(\omega\tau)^2}
\label{eq:gamma_sq}
\end{equation}
to the plate's inherent damping $\gamma_m$ and the kinetic damping $\gamma_\textrm{kin}$ discussed in Sec.~\ref{sec:free}. This squeeze film damping is, like the kinetic damping rate, proportional to the pressure.

The equation of motion for the oscillating plate amplitude can then be written as
\begin{equation}
\ddot{x}+(\gamma_m+\gamma_\textrm{kin}+\gamma_\textrm{sq})\dot{x}+(\omega^2+\kappa_\textrm{sq})x=F/m,
\label{eq:motion_single_plate}
\end{equation}
where $\kappa_\textrm{sq}=k_\textrm{sq}/m$, $m$ being the plate's mass, and $F$ is the force exerted on the plate, either by an external drive or due to thermal fluctuations.

\begin{figure}
\includegraphics[width=0.99\columnwidth]{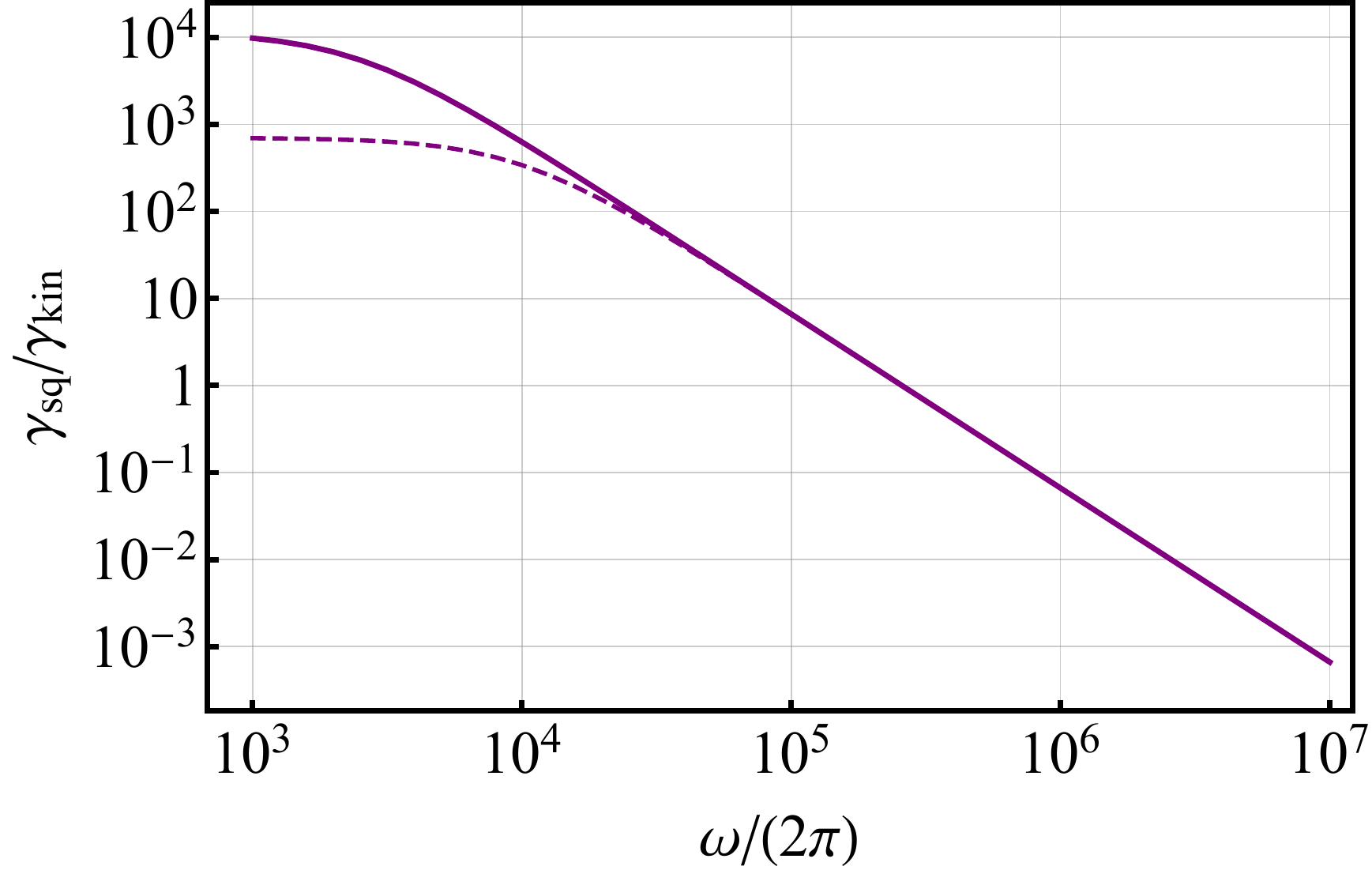}
\caption{Ratio of the squeeze film and kinetic dampings as a function of mechanical frequency $\omega_m$, at room temperature $T=298$ K and in air, for the same membrane sandwiches as in Fig.~\ref{fig:knudsen} with gaps of $d=2.1$ $\mu$m (plain) and $d=8.5$ $\mu$m (dashed).}
\label{fig:ratio}
\end{figure}

For low oscillation frequencies ($W_i\ll 1$), the squeeze film force is essentially dissipative and increases the mechanical damping by an amount proportional to the pressure, like the kinetic damping. Comparing (\ref{eq:gamma_kin}) and (\ref{eq:gamma_sq}) and for a relaxation time given by (\ref{eq:relax}), one sees that, at low Weissenberg numbers, the ratio of the squeeze film to kinetic dampings 
\begin{equation}
\frac{\gamma_\textrm{sq}}{\gamma_\textrm{kin}}\sim \frac{4}{\pi^3}\left(\frac{a}{d}\right)^2
\end{equation} 
is entirely determined by the ratio of the plate dimension to that of the gap, which can be quite large, as illustrated in Fig.~\ref{fig:ratio}.

For high oscillation frequencies ($W_i\gg 1$), the damping ratio 
\begin{equation}
\frac{\gamma_\textrm{sq}}{\gamma_\textrm{kin}}\sim \frac{\pi^3}{16}\left(\frac{\bar{v}}{a\omega}\right)^2
\end{equation} 
becomes independent of the gap size and decreases as $1/\omega^2$ (Fig.~\ref{fig:ratio}). The squeeze film force is then essentially elastic and results in an increased stiffness for the plate. In this limit, the gas-added spring constant is thus directly proportional to the pressure and results in a frequency shift of the plate mechanical resonance frequency given by the simple expression
\begin{equation}
\tilde{\omega}^2=\omega^2+\kappa_{sq}=\omega^2+\frac{P}{\rho t d},
\end{equation}
where $\rho$ is the plate density and $t$ its thickness.

\subsection{Squeeze film effects in membrane sandwiches}

We now consider the vibrations of two nearly-identical clamped, square membranes, parallel to each other and separated by a small gap $d\ll a$, as illustrated in Fig.~\ref{fig:membranes}(c). We denote by $x_1$ and $x_2$ the (small) amplitudes of two drummodes with intrinsic (in vacuum) frequencies $\omega_1$ and $\omega_2$ and damping rates $\gamma_1$ and $\gamma_2$. We assume that the membranes are nearly-identical and possess high mechanical quality factors $Q_i=\omega_i/\gamma_i$ ($i=1,2$), so that the two drummodes considered have similar intrinsic frequencies ($|\omega_1-\omega_2|\ll\omega_1,\omega_2$), although these might still differ by many intrinsic linewidths. The compression of the gas between the two oscillating membranes results in opposite squeeze film forces on each of them. Their derivation proceeds indeed from Eq.~({\ref{eq:Deltan}), replacing $x$ by the relative displacement between the membranes, $x_1-x_2$. Consequently, the amplitudes of both modes become dynamically coupled and their dynamics are dictated by the following generalizations of (\ref{eq:motion_single_plate})
\begin{align}
\ddot{x}_1+(\gamma_1+\gamma_\textrm{kin}+\gamma_{\textrm{sq},1})\dot{x}_1+\omega_1^2x_1+\kappa_\textrm{sq}(x_1-x_2)=\frac{F_1}{m_1},\label{eq:motion_x1}\\
\ddot{x}_2+(\gamma_2+\gamma_\textrm{kin}+\gamma_{\textrm{sq},2})\dot{x}_2+\omega_2^2x_2+\kappa_\textrm{sq}(x_2-x_1)=\frac{F_2}{m_2}.\label{eq:motion_x2}
\end{align}
Let us note that these equations are valid as long as the two frequencies of the two drummodes considered are well-separated from the other mode frequencies, so that their off-resonant contributions are negligible at the frequencies considered. This is indeed justified as the dynamics of the coupled modes depend on the amplitude of the pressure-induced coupling as compared to the difference in their intrinsic frequencies. In this two-mode picture, very non-degenerate frequency modes will oscillate independently of each other, their respective mechanical frequencies being positively shifted by the same amount as the pressure is increased. In contrast, nearly degenerate frequency modes will hybridize into a bright mode (corresponding to the relative motion $x_1-x_2$) and a dark mode (corresponding to the center-of-mass mode $x_1+x_2$), as the pressure-induced coupling is increased.

Fourier transforming Eqs.~(\ref{eq:motion_x1},\ref{eq:motion_x2}) and assuming that the mechanical quality factors remain high over the pressure range considered ($\omega_i\gg\gamma_i+\gamma_\textrm{kin}+\gamma_{\textrm{sq},i}$), the frequencies of the bright and dark modes, $\omega_+$ and $\omega_-$, respectively, are straightforwardly found to be given by
\begin{equation}
\omega_{\pm}=\left[\omega_0^2+\delta^2+2\eta\omega_0\pm 2\omega_0\sqrt{\delta^2+\eta^2}\right]^{1/2},
\end{equation}
 where
\begin{equation}
\omega_0=\frac{\omega_1+\omega_2}{2},\hspace{0.2cm}\delta=\frac{\omega_1-\omega_2}{2},\hspace{0.2cm}\textrm{and}\hspace{0.2cm}\eta=\frac{\kappa_\textrm{air}}{2\omega_0}.
\end{equation}

\begin{figure}
\includegraphics[width=0.9\columnwidth]{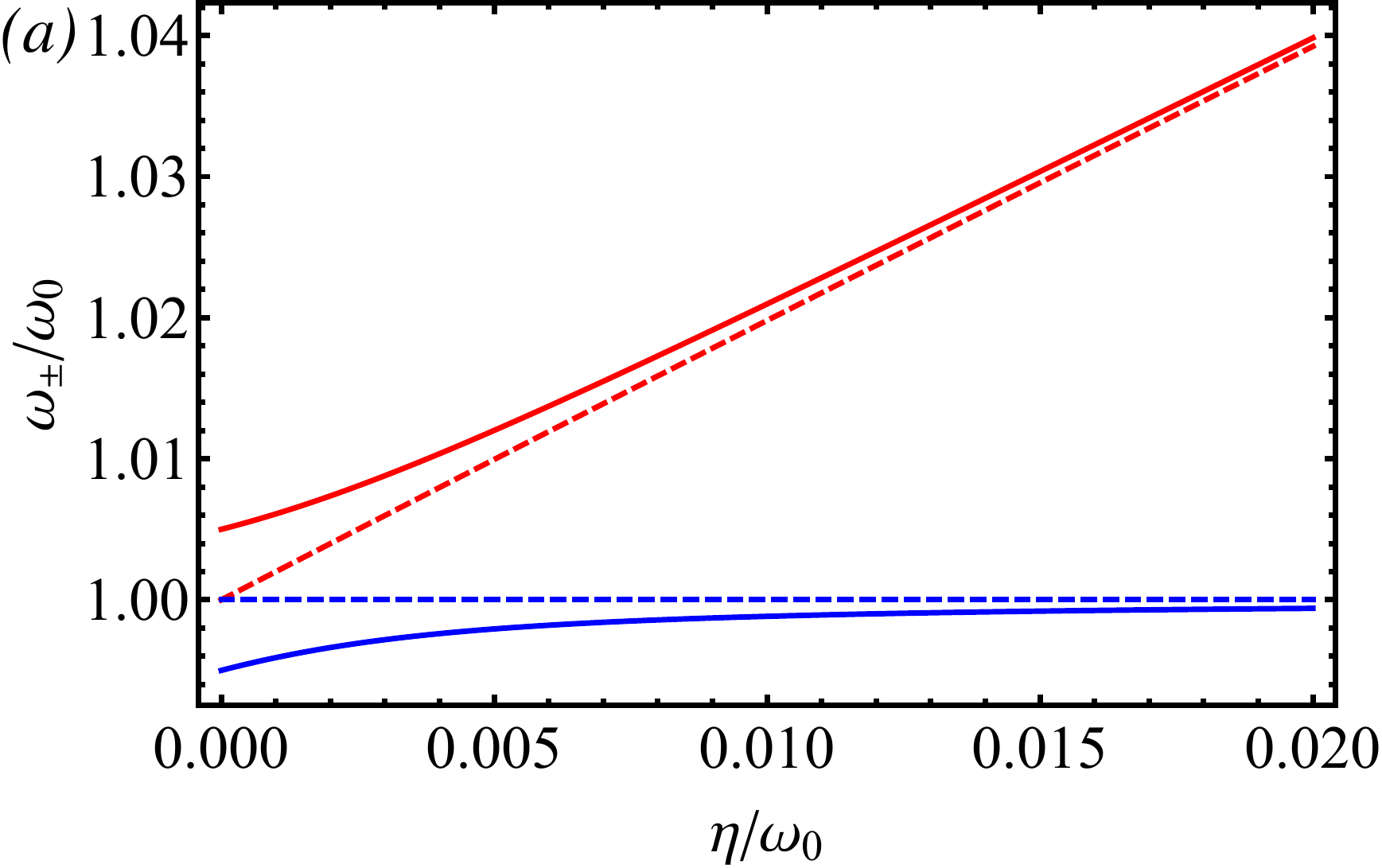}
\llap{\parbox[b]{6.1in}{\large{{\it (a)}}\\\rule{0ex}{1.8in}}}
\includegraphics[width=0.88\columnwidth]{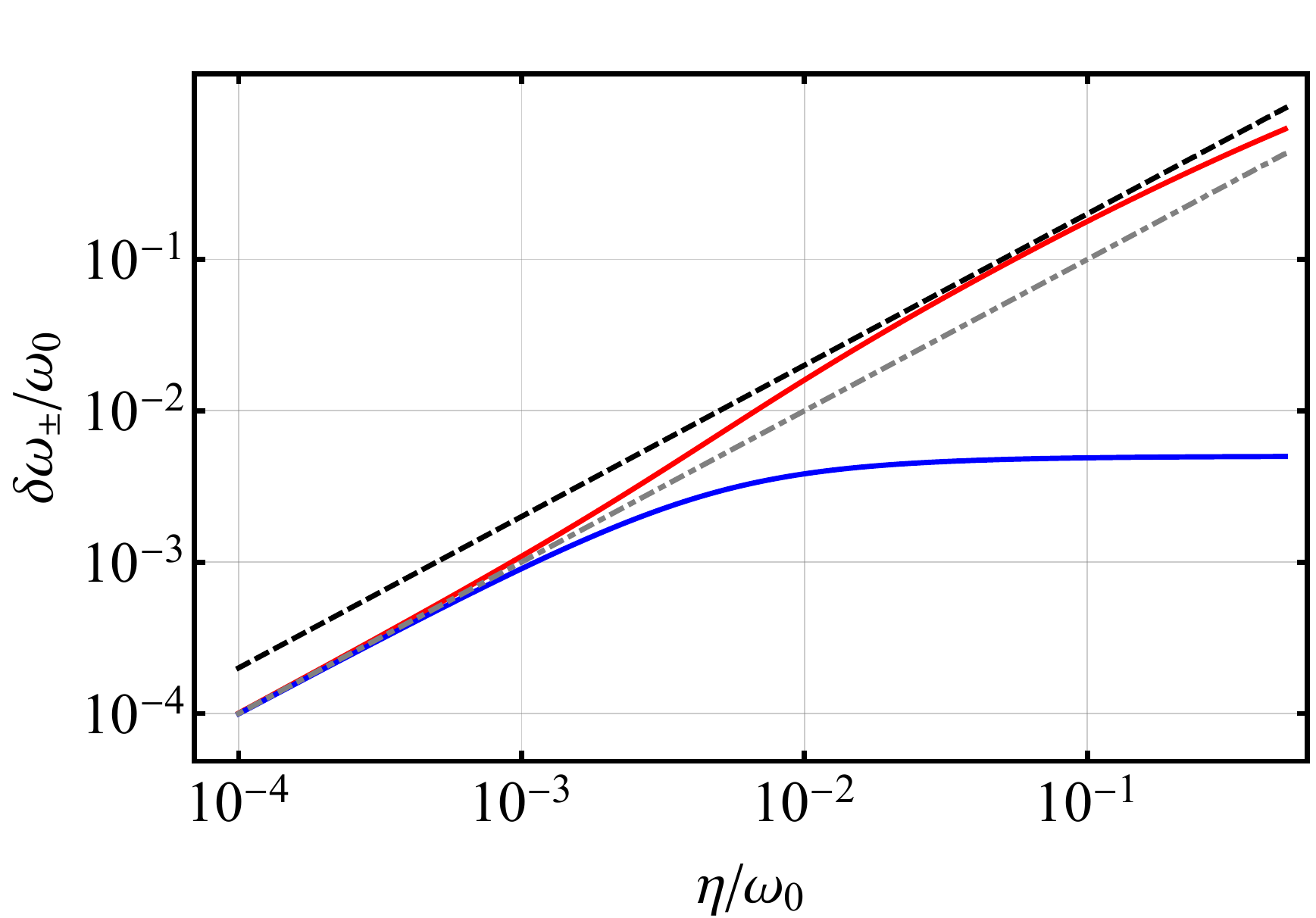}
\llap{\parbox[b]{6.1in}{\large{{\it (b)}}\\\rule{0ex}{1.8in}}}
\caption{(a) Bright (red) and dark (blue) mode frequencies $\omega_{\pm}$ as a function of gas-induced coupling $\eta$ when $\delta=0$ (dashed lines) and $\delta=0.005\omega_0$ (full lines). (b) Bright and dark mode frequency shifts $\delta\omega_{\pm}$ as a function of $\eta$ (log-log scale) for $\delta=0.005\omega_0$. The dot-dashed gray and the dashed black lines show linear shifts equal to $\eta$ and $2\eta$, respectively. Figure adapted from~\cite{Naserbakht2019sna}.}
\label{fig:shifts}
\end{figure}

Figure~\ref{fig:shifts} illustrates the variations of the bright and dark mode frequencies, $\omega_\pm$, as the gas-induced coupling $\eta$ is increased. When the membranes' intrinsic degenerate frequencies are perfectly degenerate ($\delta=0$), the dark mode frequency remains unchanged, while that of the bright mode experiences a linear shift given by $2\eta$. For nondegenerate intrinsic frequencies, positive linear frequency shifts given by $\eta$ are first observed as long as $\eta\ll\delta$. When $\eta\sim\delta$, the dark mode becomes decoupled from the pressure and its frequency converges toward $\omega_0$, while the bright mode experiences a doubled linear shift given by $2\eta$. When $\eta$ becomes so large that the frequency shift is no longer small with respect to the bare frequency, the frequency shift no longer scales linearly with the coupling, but as $\sqrt{2\eta}$), as illustrated in Fig.~\ref{fig:shifts}b.

\section{Experimental realization with SiN membrane sandwiches}

\subsection{Fabrication and measurement procedure}

\begin{figure}
\includegraphics[width=0.9\columnwidth]{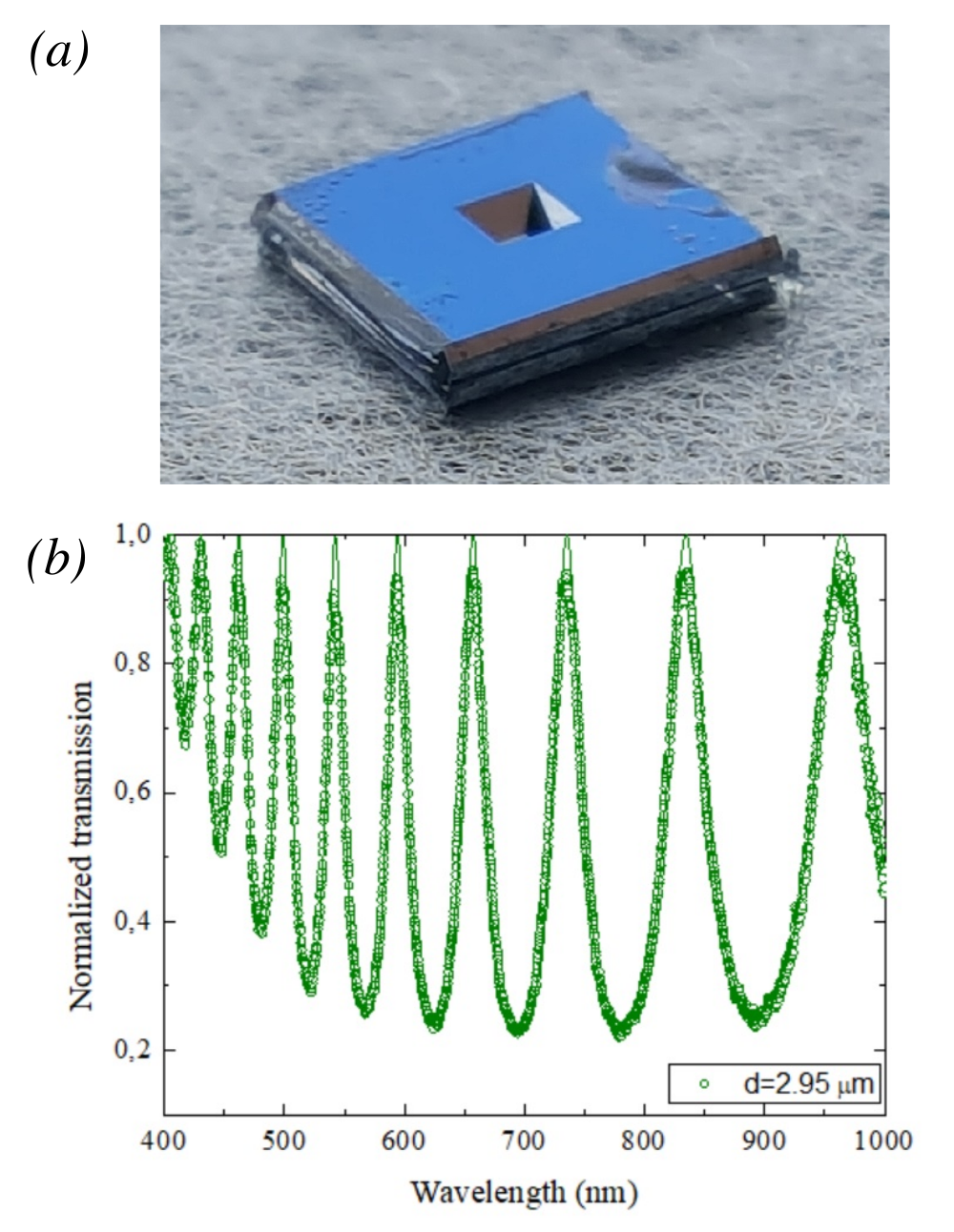}
\caption{(a) Picture of a 2.95 $\mu$m-gap membrane sandwich. (b) Normalized transmission spectrum of the membrane sandwich under broadband illumination. The solid line shows the results of a fit with a linear Fabry-Perot transmission model. Figure adapted from~\cite{Naserbakht2019sna}.}
\label{fig:trans}
\end{figure}

The membrane resonators used in the examples discussed below are commercial (Norcada Inc.), high tensile stress ($\sim 0.7-0.9$ GPa), 500 $\mu$m-square and $\sim 100$ nm-thick stoichiometric silicon nitride films deposited on a 5 mm-square and 500 $\mu$m-thick silicon frame by low pressure chemical vapor deposition. The membrane sandwiches used in this work were realized using the method described in~\cite{Nair2017}: a spacer was deposited on one of the SiN films and the membranes were positioned parallel to each other using actuators and monitoring the transmission of broadband light through the Fabry-Perot interferometer constituted by both membranes. A picture of an assembled sandwich with a gap of $d=2.95$ $\mu$m and its normalized transmission spectrum are shown in Fig.~\ref{fig:trans}. The membrane sandwiches are then placed in a vacuum chamber, and positioned parallel with a 50:50 beamsplitter mirror to form a 7 mm-long Fabry-Perot interferometer, whose length can be controlled by means of a piezoelectric transducer (Fig.~\ref{fig:setup}). Monochromatic light from a tunable external cavity diode laser (890-940 nm) is injected into the interferometer and the transmitted light is detected by a fast photodiode and its fluctuations recorded on a narrow resolution bandwidth spectrum analyzer. The length of the interferometer is chosen so as to maximize the displacement sensitivity. The recorded thermal noise spectra around the mechanical resonance frequencies allow for their precise determination, as well as that of the mechanical quality factors. The pressure is varied by letting air at room temperature in the chamber.

The fundamental drummodes of the membranes typically have intrinsic (in vacuum) resonance frequencies between 700 and 850 kHz and display quality factors of the order $10^5$ after assembly. Typical variations of the Knudsen number with pressure in the range investigated are shown in Fig.~\ref{fig:knudsen}, illustrating that the transition regime occurs in the few to few tens of millibars range. For sanwiches with gaps between 2 and 8.5 microns and the frequencies considered, the Weissenberg number is $\sim 80-340$, so that the resonators operate well in the high frequency regime, in which the squeeze film force is expected to be essentially elastic. This behavior is also expected in the transition and viscous regimes, since estimates of the squeeze parameter also give high values, as can be seen on Fig.~\ref{fig:knudsen}. Figure~\ref{fig:ratio} additionally shows that, for the high mechanical frequencies considered, squeeze film damping is expected to be negligible with respect to kinetic damping.

\begin{figure}
\includegraphics[width=0.85\columnwidth]{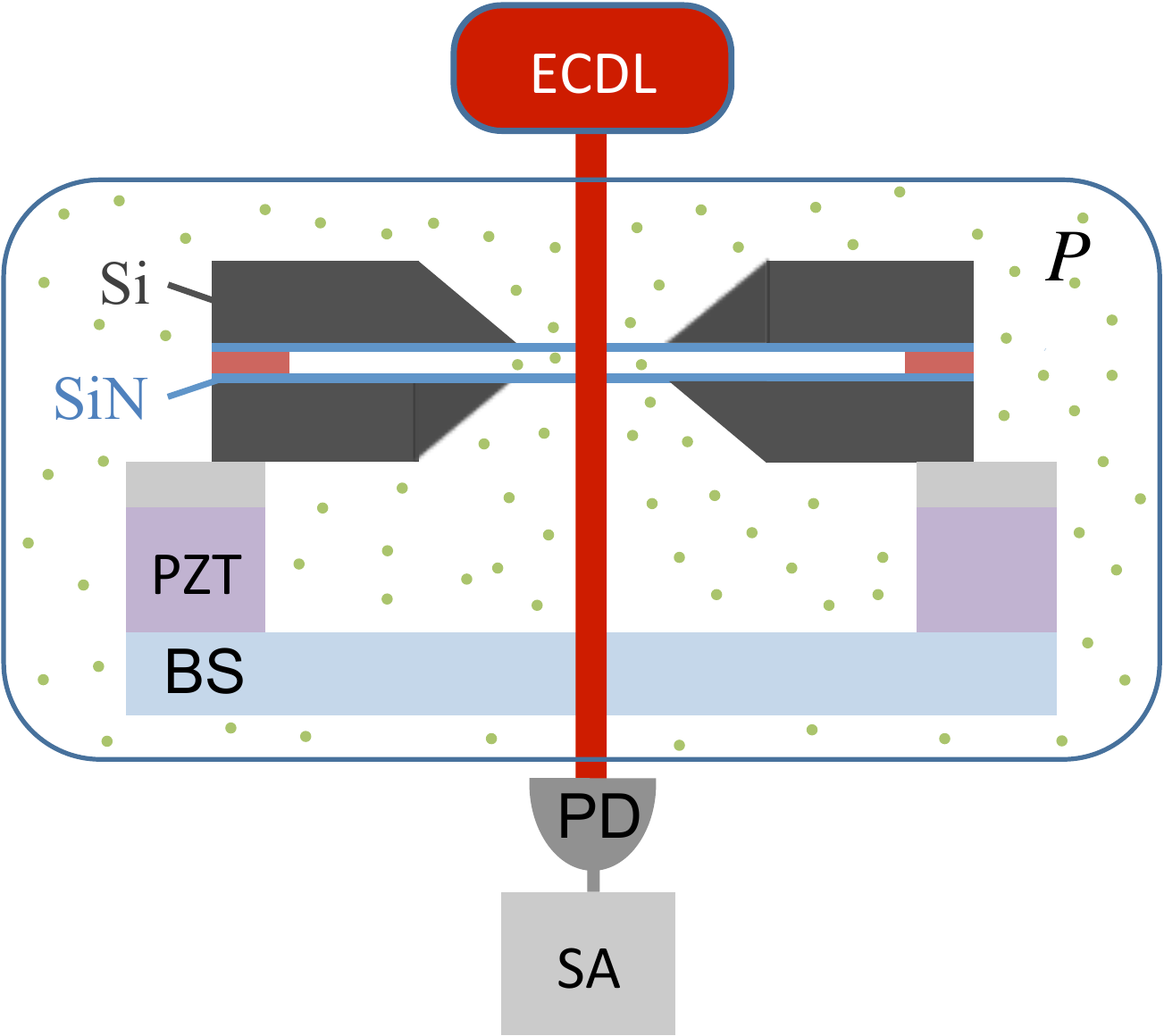}
\caption{Schematic of the experimental setup used for investigating the effects of pressure on membrane sandwiches. ECDL: external cavity diode laser. PZT: piezoelectric transducer. BS: beamsplitter. PD: photodiode. SA: spectrum analyzer. Figure adapted from~\cite{Naserbakht2019sna}.}
\label{fig:setup}
\end{figure}

\subsection{8-9 $\mu$m-gap sandwiches}

\begin{figure}
\includegraphics[width=0.9\columnwidth]{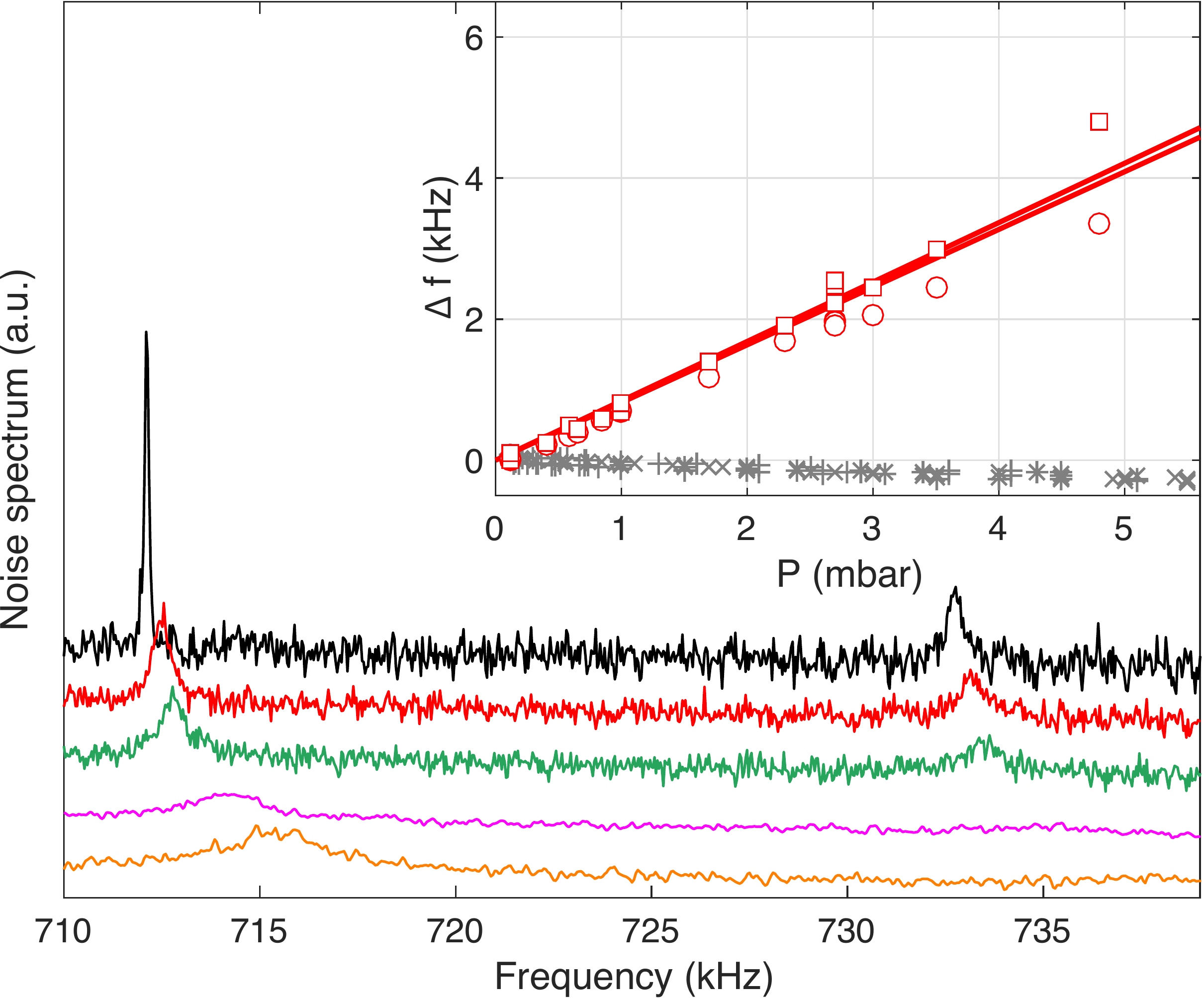}
\llap{\parbox[b]{6.3in}{\large{{\it (a)}}\\\rule{0ex}{2.4in}}}\\
\vspace{0.2cm}
\includegraphics[width=0.9\columnwidth]{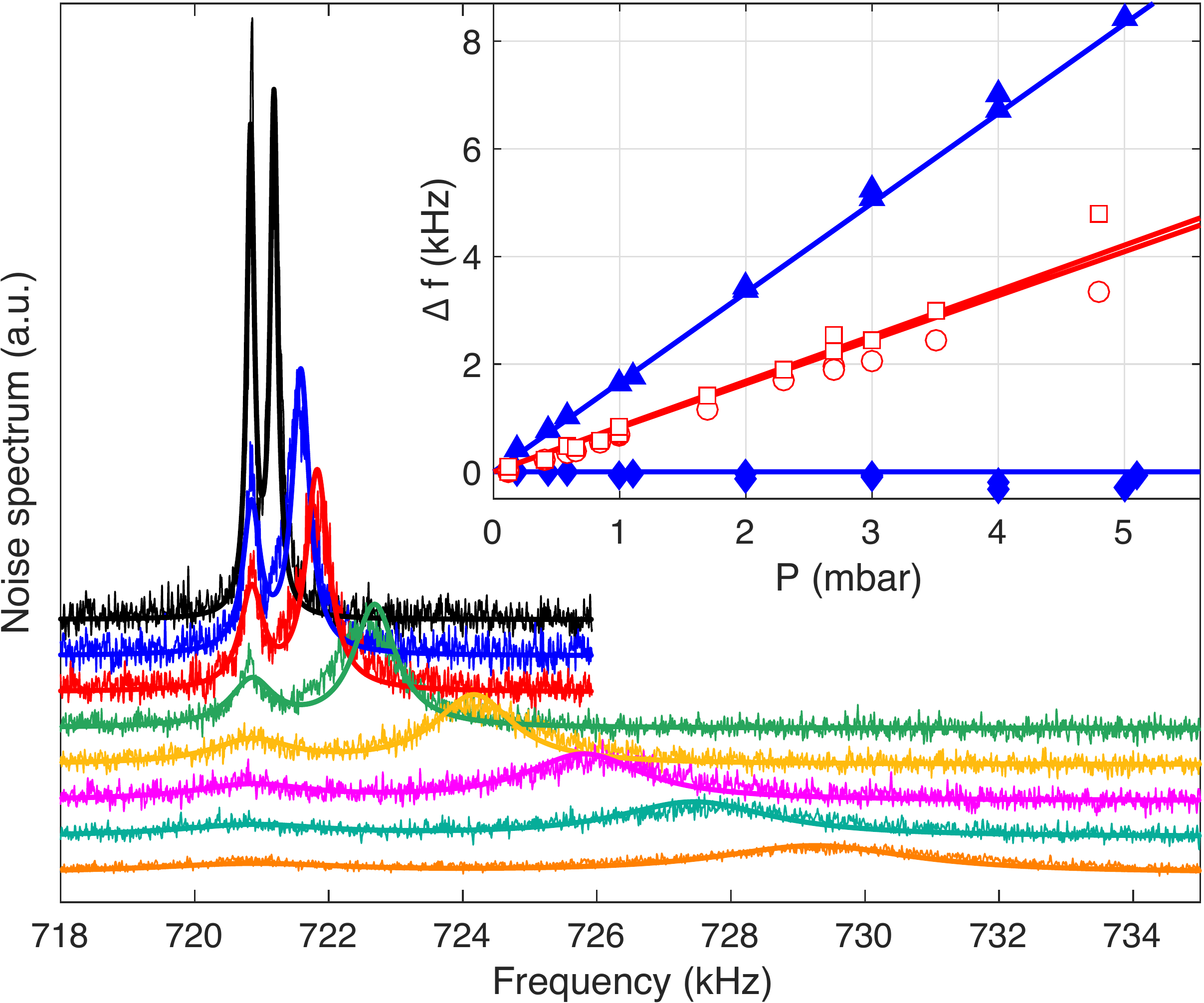}
\llap{\parbox[b]{6.3in}{\large{{\it (b)}}\\\rule{0ex}{2.4in}}}
\caption{(a) Thermal noise spectra around the fundamental drummode resonance frequencies of a {\it non-degenerate} 8.5 $\mu$m-gap sandwich for different pressures in the 0.1-5 mbar range (the spectra are vertically offset for clarity). The inset shows the resonance frequency shifts for both modes (red dots and squares). The red lines show the squeeze film model predictions and the gray crosses show the (negative) resonance frequency shift of a single membrane resonator. (b) Thermal noise spectra around the fundamental drummode resonance frequencies of a {\it near-degenerate} 8.5 $\mu$m-gap sandwich for different pressures in the same range. The solid lines show the results of fits to the coupled oscillator model based on the elastic squeeze film effect predictions. The inset show the resonance frequency shifts of the bright and dark modes (blue triangles and losanges), the blue lines showing the squeeze film model predictions. The red points and lines show again the data of the inset of (a) for visual comparison. Figure reproduced from~\cite{Naesby2017}.}
\label{fig:medium}
\end{figure}

First investigations were carried out with membrane sandwiches having gaps in the $8-9$ $\mu$m, provided by a predeposited dielectric spacer on one of the commercial chips~\cite{Nair2017}. Thermal noise spectra around the fundamental drummode frequencies were recorded at various air pressure at room temperature and for different sandwiches~\cite{Naesby2017}. Figure~\ref{fig:medium}(a) shows examples of such spectra in the few millibar range for a {\it non-degenerate} sandwich for which the intrinsic frequencies of the fundamental modes were 712 and 733 kHz, corresponding to $\delta=0.014\omega_0$. Positive linear resonance frequency shifts of 0.84 and 0.82 kHz/mbar, respectively, are observed in good agreement with the elastic squeeze film predictions. In contrast, measurements performed with a single membrane resonator from the same fabrication batch showed comparatively much smaller, negative resonance frequency shits (inset of Fig.~\ref{fig:medium}(a)).

The {\it near-degenerate} sandwich case is shown in Fig.~\ref{fig:medium}(b), where a clear air-induced hybridization of the fundamental drummodes with intrinsic frequencies of 721 and 721.2 kHz ($\delta=10^{-4}\omega_0$) can be observed. There, positive resonance frequency shifts of $\simeq 0$ and 1.66 kHz/mbar are observed for the bright and dark modes, respectively, again in good agreement with the theoretical predictions. The solid lines in Fig.~\ref{fig:medium}(b) also show the results of fits to the theoretically expected thermal noise spectra of coupled harmonic oscillators, in which $\gamma_\textrm{kin}$ and $\kappa_{sq}$ are given by the theoretical predictions based on the independently determined membrane thickness and gap. The only free fitting parameters are a global scaling factor for the thermal noise spectrum amplitude and the relative weight of the measured membrane displacements, which non-trivially depend on a number of parameters (interferometer length, intermembrane separation, optical/mechanical mode overlap, membrane parallelism, etc.). A very good agreement is observed between the measured spectra and the coupled oscillator model predictions.

\begin{figure}
\includegraphics[width=0.9\columnwidth]{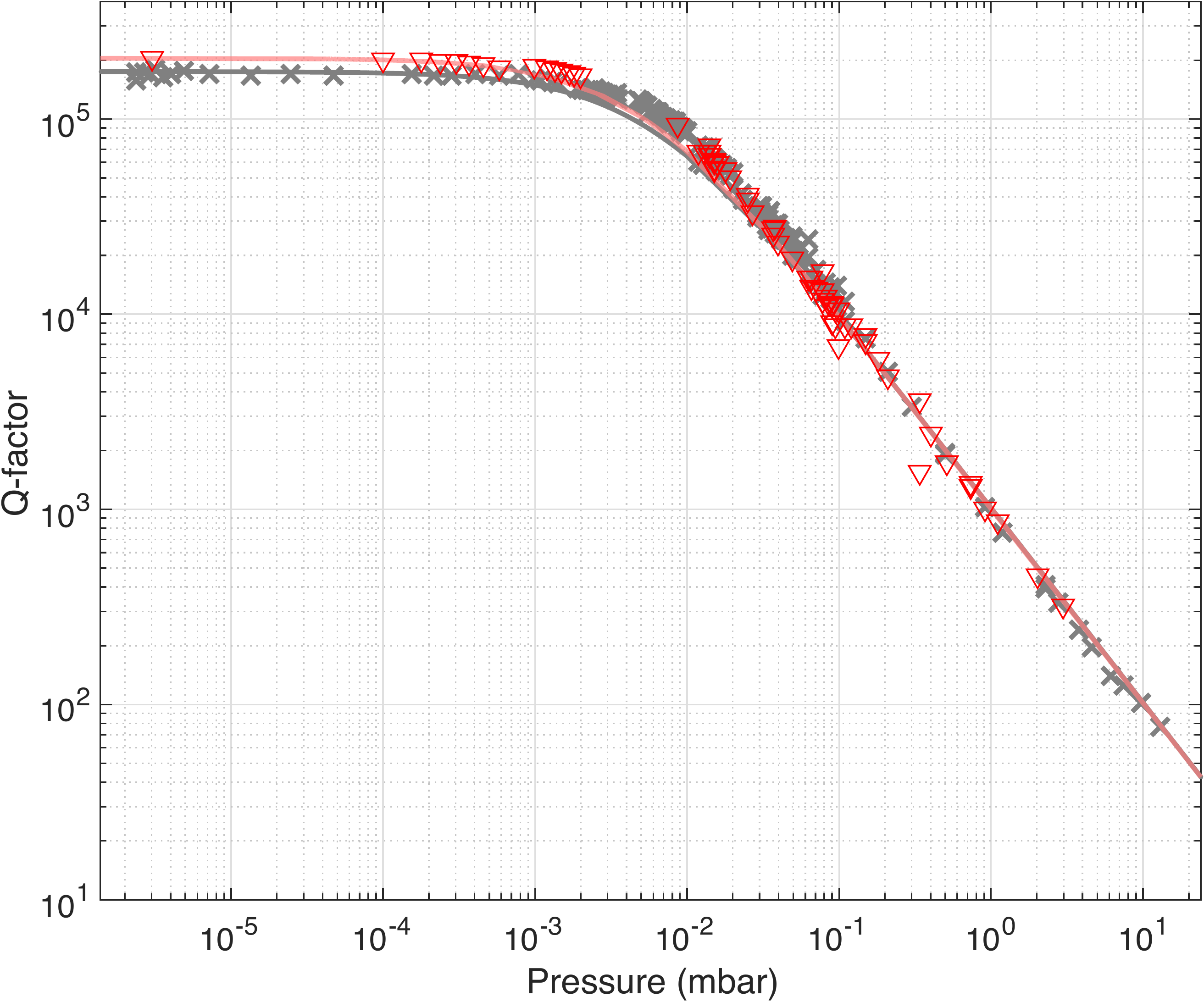}
\caption{Variations with pressure of the mechanical quality factor of the fundamental drummode of a single membrane (gray crosses) and of a membrane in a 8.5 $\mu$m-gap membrane sandwich (red triangles). The solid lines show the kinetic damping predictions. Figure reproduced from~\cite{Naesby2017}.}
\label{fig:mediumQ}
\end{figure}

To further ascertain that the squeeze film damping contribution was indeed negligible with respect to that of the kinetic damping, the variations with the pressure of the quality factor of the fundamental mode of a single free-standing membrane and of a membrane in an 8.5 $\mu$m-gap sandwich were accurately measured in the range $10^{-6}-10$ mbar and are displayed in Fig.~\ref{fig:mediumQ}. The same linear increase of the quality factor is observed as the pressure is reduced, until air damping becomes negligible with respect to the intrinsic damping at pressures below  $10^{-3}$ mbar. The measured quality factors were also found to be in very good agreement with the kinetic damping predictions of Eq.~(\ref{eq:gamma_kin}).

\subsection{2-3 $\mu$m-gap sandwiches}

\begin{figure}
\includegraphics[width=0.9\columnwidth]{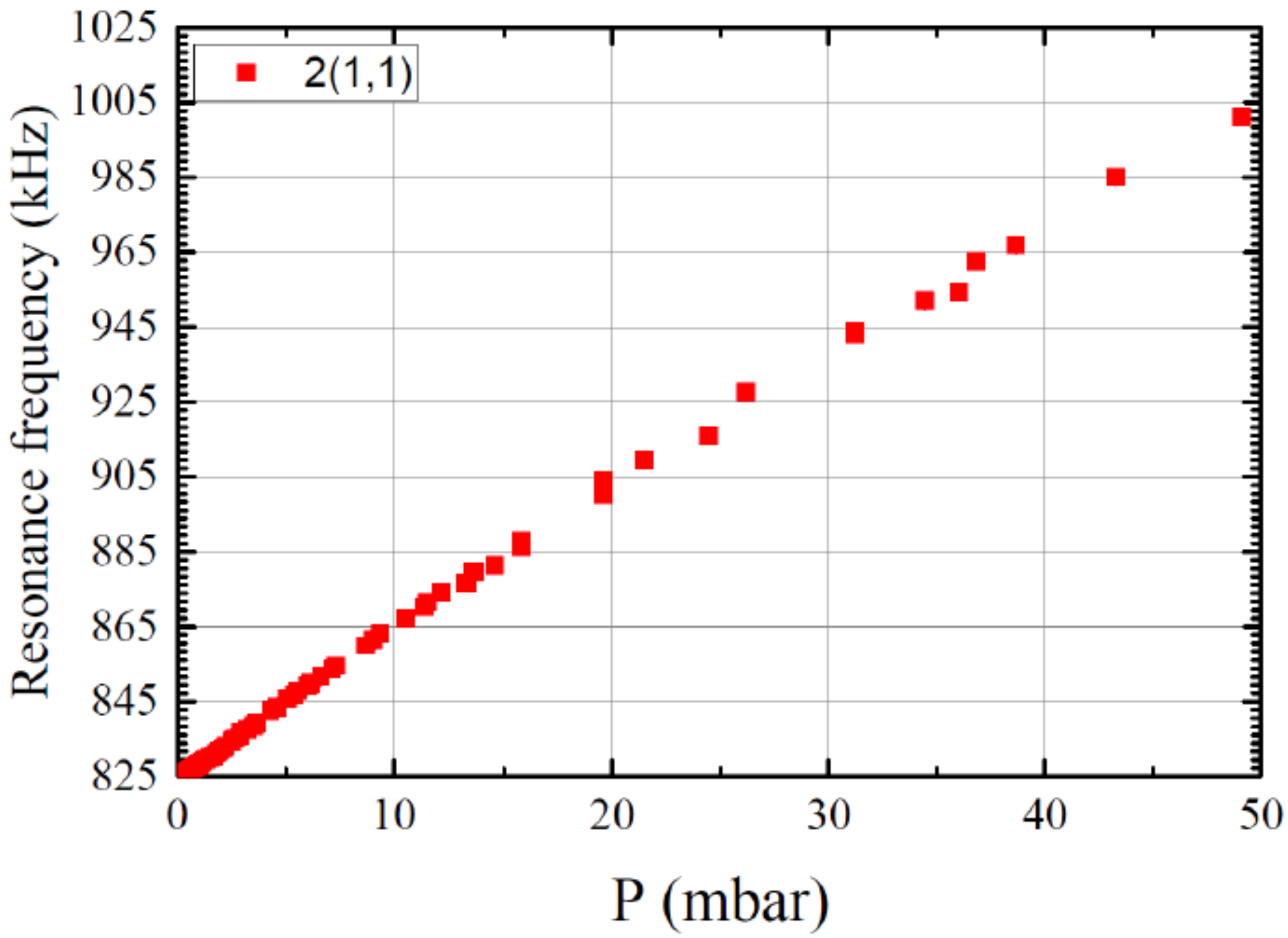}
\llap{\parbox[b]{6.2in}{\large{{\it (a)}}\\\rule{0ex}{2.0in}}}\\
\vspace{0.2cm}
\includegraphics[width=0.9\columnwidth]{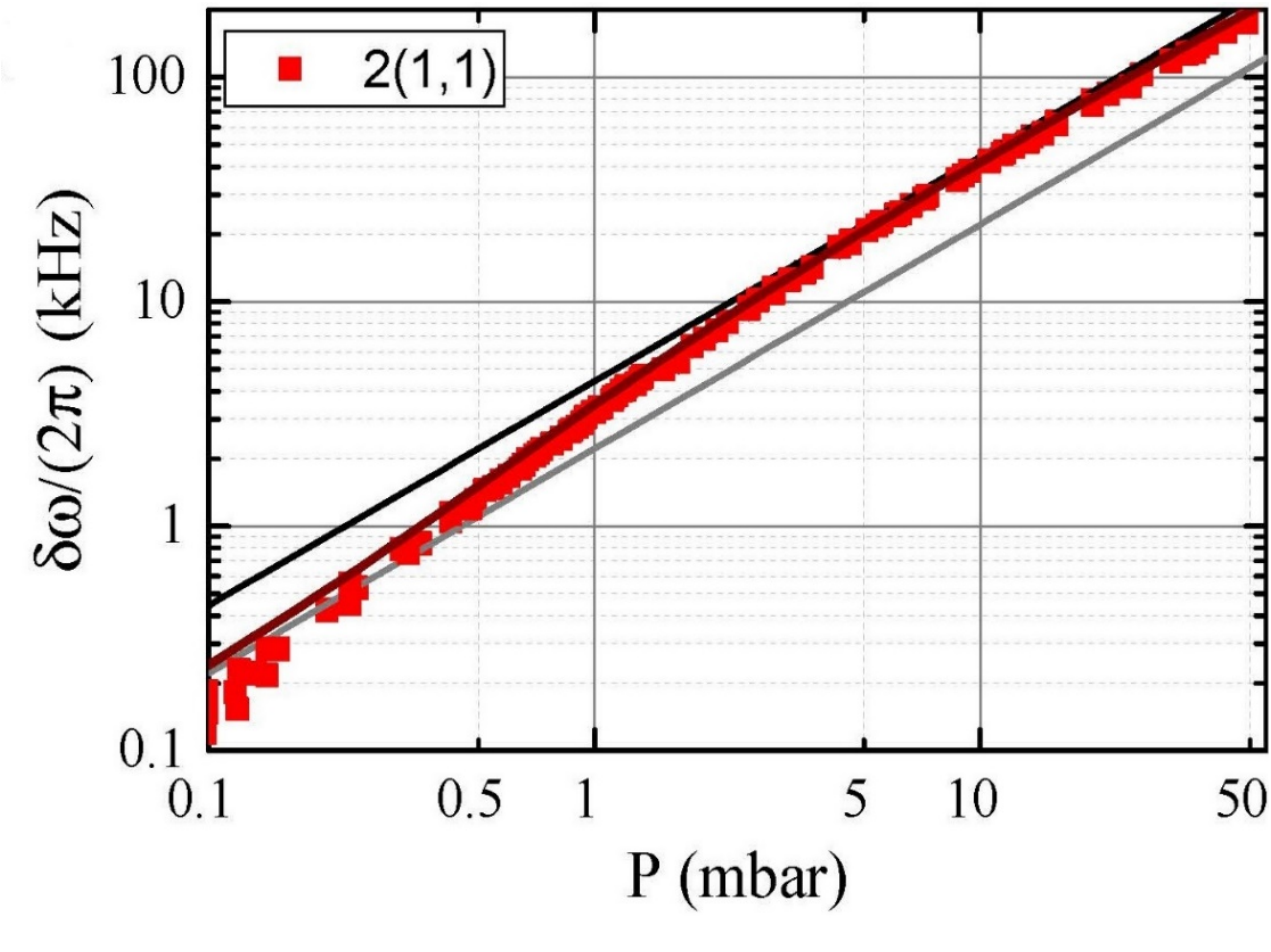}
\llap{\parbox[b]{6.1in}{\large{{\it (b)}}\\\rule{0ex}{2.1in}}}
\caption{(a) Resonance frequency and (b) frequency shift (log-log scale) of the fundamental drummode of a membrane in a 2.95 $\mu$m-gap sandwich, as a function of air pressure. The solid dark red line in (b) shows the predictions of the squeeze film model, and the single and double linear shifts illustrated in Fig.~\ref{fig:shifts}(b) are shown by the gray and black lines, respectively. Figure adapted/reproduced from~\cite{Naserbakht2019sna}.}
\label{fig:Al}
\end{figure}

To increase the magnitude of the squeeze film effects, sandwiches with shorter gaps (2-3 $/mu$m) were assembled using similar membranes, albeit with slightly lower ($\sim 87$~nm) thickness and higher ($\sim 0.9$~GPa) tensile stress, and various spacers made of aluminium or UV-light cured resist. Figure~\ref{fig:Al} shows the effects of pressure on the fundamental drummode resonance frequency of a 2.95 $\mu$m-gap sandwich in the range $0.1-50$ mbar. Very large squeeze film-induced shifts of up to 175 kHz---representing a relative variation of $\sim 20\%$ of the fundamental drummode frequency---are observed at 50 mbar. In the range 2-20 mbar a pressure responsitivity of about 4 kHz/mbar is achieved, which is a factor approximately 2 lower than the largest squeeze film sensor responsitivities reported with graphene microdrums~\cite{Dolleman2016}. The crossovers from the linear, independent membrane shift to the doubled linear, coupled membrane shift---and even the nonlinear strong coupling shift discussed in the previous section---are clearly observed. The measured shifts also match very well the theoretical expectations based on the independently determined intermembrane gap and membrane thickness. The variations of frequency shifts and damping rates were furthermore measured for higher order modes and found to evolve as well as expected from the squeeze film model predictions~\cite{Naserbakht2019sna}.

To evaluate the pressure sensitivity of these sensors at low pressure, systematic measurements of the resonance frequencies referenced to that in vacuum were performed by sequential rapid increase and decrease of the pressure in order to minimize the thermal drifts during the measurements. For these measurements, a 2.1 $\mu$m-gap sandwich with membranes having intrinsic fundamental drummode resonance frequencies of 820.1 and 831.2 kHz with quality factors 86000 and 46000, respectively, was used. At low pressure, an equal pressure responsitivity of 3.1 kHz/mbar at low pressures is observed for each mode with a sensitivity at the $\mu$bar level, essentially limited by the thermal drifts of the vacuum chamber used in these measurements.

\begin{figure}
\includegraphics[width=\columnwidth]{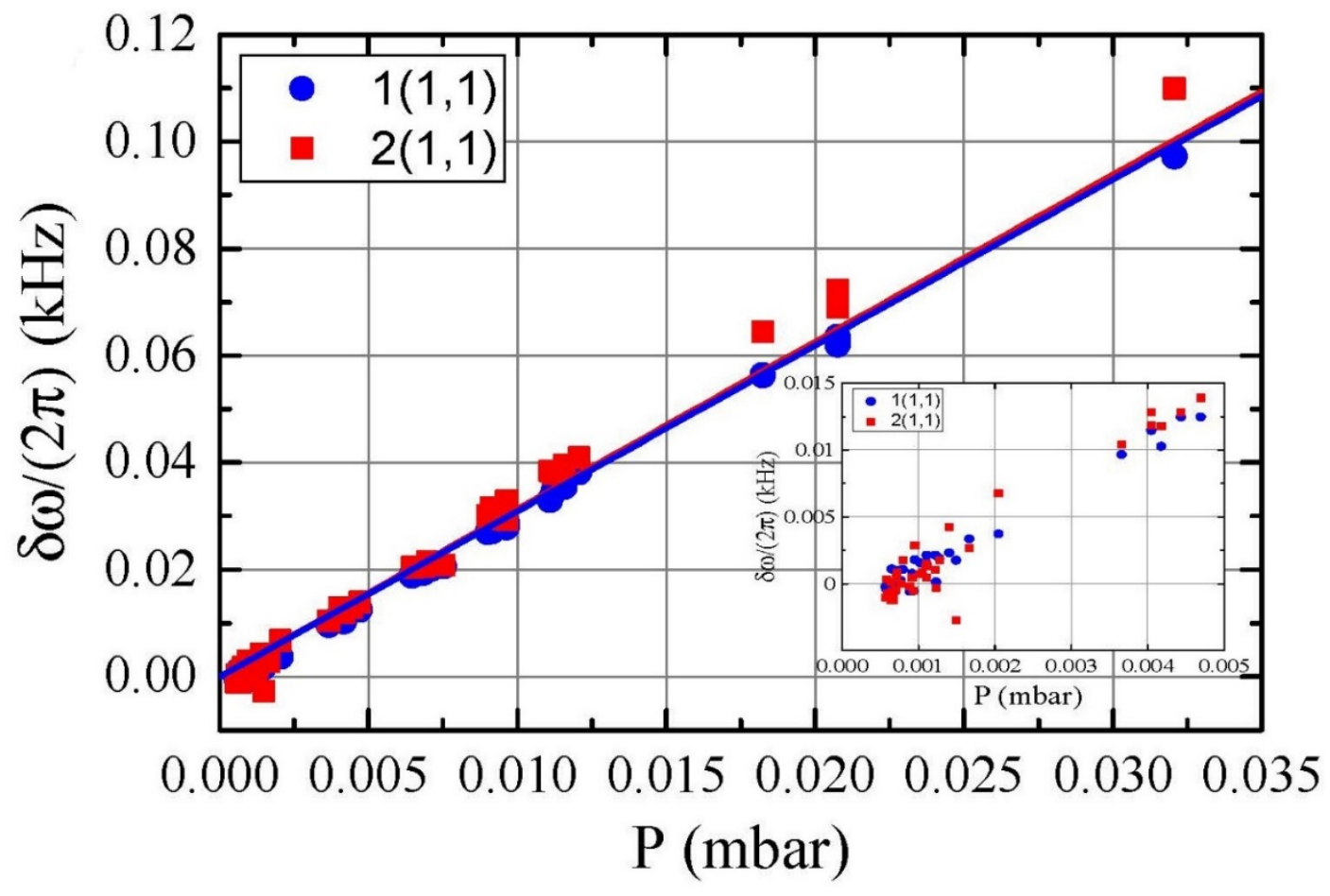}
\caption{Resonance frequency shifts of the fundamental drummodes of a 2.1 $\mu$m-gap sandwich as a function of pressure. The circles and dots represent the experimental data, the lines the theoretical predictions and the inset shows a zoom-in onto the few $\mu$bar region. Figure reproduced from~\cite{Naserbakht2019sna}.}
\label{fig:S10}
\end{figure}

\section{Discussion}

The workings of membrane sandwich pressure sensors have been introduced, with a particular focus on the free molecular flow regime, and recent experimental implementations using silicon nitride resonators operating in the high frequency regime have been presented. 

The use of a membrane sandwich in which a fluid is isothermally compressed in the small gap region between the membranes makes it possible to exploit squeeze film effects in order to determine in a direct and species-independent way the pressure exerted on the membranes. High pressure responsivity/sensitivity can then be obtained, as small gaps (few microns) and thin membranes (<100nm) make for strong squeeze film effects, high tensile stress make for high frequency ($\sim$ MHz) and high mechanical quality factor modes, which, combined with the resonators' large area, ensures effective trapping of the gas. The gas-induced coupling between the modes is also shown to enhance the pressure responsivity. In addition, the "canonical" large, unperforated, thin, parallel membrane geometry allows for a direct comparison with the theoretical predictions.

Interesting prospects for improving the sensitivity of such membrane sandwich pressure sensors at low pressures exist e.g. via dissipation dilution and nanostructuring to dramatically improve their mechanical quality factors~\cite{Tsaturyan2017}. Extending the sensitivity of these devices further into the high vacuum or ultrahigh vacuum regime would be relevant for a wide range of applications, ranging from absolute high vacuum pressure calibration~\cite{Volklein2007,Gorecka2009} to the direct determination of the vapor pressure of chemically or environmentally relevant low volatile substances~\cite{Bilde2015}. Let us note that, although this article focused primarily on the free molecular flow regime, similar concepts could be applied to the viscous regime, which could be explored e.g. with higher frequency modes. 

Additionally, improving the optical reflectivity of the suspended films by patterning of nanostructures~\cite{Kemiktarak2012,Norte2016,Reinhardt2016,Moura2018,Naesby2018,Nair2019} or exploiting the squeeze film-induced coupling could also be exploited via e.g. electrical tuning of the mechanics~\cite{Naserbakht2019apl,Naserbakht2019appsci} also represent interesting avenues to explore for improving the sensitivity of such sandwich sensors.

\begin{acknowledgments}
The author is grateful to Andreas Naesby and Sepideh Naserbakht for their contributions to this work and acknowledges financial support from Villumfonden and Independent Research Fund Denmark. The data that support the findings of this study are available from the corresponding author upon request.
\end{acknowledgments}

%\appendix

%\section{Appendixes}

%To start the appendixes, use the \verb+\appendix+ command.
%This signals that all following section commands refer to appendixes
%instead of regular sections. Therefore, the \verb+\appendix+ command
%should be used only once---to set up the section commands to act as
%appendixes. Thereafter normal section commands are used. The heading
%for a section can be left empty. For example,
%\begin{verbatim}
%\appendix
%\section{}
%\end{verbatim}
%will produce an appendix heading that says ``APPENDIX A'' and
%\begin{verbatim}
%\appendix
%\section{Background}
%\end{verbatim}
%will produce an appendix heading that says ``APPENDIX A: BACKGROUND''
%(note that the colon is set automatically).
%
%If there is only one appendix, then the letter ``A'' should not
%appear. This is suppressed by using the star version of the appendix
%command (\verb+\appendix*+ in the place of \verb+\appendix+).

\nocite{*}
\bibliography{pressure_sensor_bib.bib}

\end{document}